# Title: Exoplanet Radius Gap Dependence on Host Star Type

**Authors:** Li Zeng*[1], Stein B. Jacobsen[1], Dimitar D. Sasselov[2]

**Affiliations:**
[1]Dept of Earth & Planetary Sciences, Harvard University, 20 Oxford St., Cambridge, MA02138
[2]Harvard-Smithsonian Center for Astrophysics, 60 Garden St., Cambridge, MA02138
Correspondence to: astrozeng@gmail.com

**Introduction:**
Exoplanets smaller than Neptune are numerous, but the nature of the planet populations in the 1-4 Earth radii ($R_\oplus$) range remains a mystery. The complete *Kepler* sample of Q1-Q17 exoplanet candidates shows a radius gap at ~ 2 $R_\oplus$, as reported by us in January 2017 in *LPSC* conference abstract #1576 (Zeng et al. 2017). A careful analysis of *Kepler* host stars spectroscopy by the CKS survey allowed Fulton et al. (2017) in March 2017 to unambiguously show this radius gap. The cause of this gap is still under discussion (Ginzburg et al. 2017; Lehmer & Catling 2017; Owen & Wu 2017). Here we add to our original analysis the dependence of the radius gap on host star type.

**Result:**
We have found that the exoplanets' radius gap depends on the host star type and varies from 1.3 to 2.2 $R_\oplus$. **Figure 1** shows our detailed analysis of 4433 *Kepler* planet candidates from Q1-Q17 (*NASA* Exoplanet Archive (Akeson et al. 2013)):

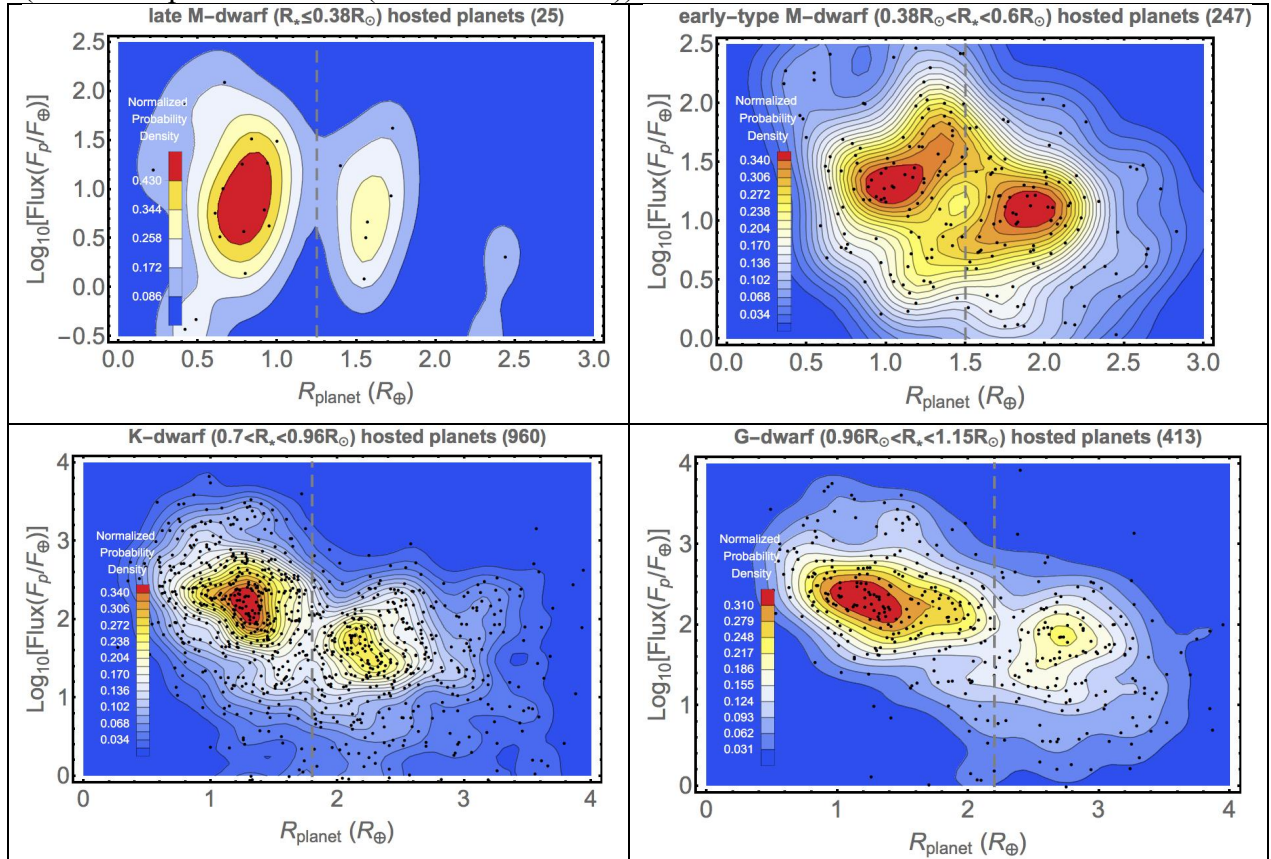

**Figure 1.** Comparison of 2-dimensional histograms of irradiance-versus-radius distribution for late-type M-dwarf, early-type M-dwarf, K-dwarf, and G-dwarf hosted planets in the *Kepler* sample (main-sequence stars only). The number of planets in each plot is shown in parentheses. It shows that the bi-modal distribution also applies to M-dwarf hosted planets, but with a gap at somewhat smaller radius. The distributions on either side of the gap have significant scatter and overlap of at least one-order-of-magnitude in the stellar irradiance.

**Discussion:**

The radius gap ($R_{gap}$) does seem to depend on the host star type, especially for host stars significantly smaller than the Sun. Although we do not have enough statistics for the extremely small late-type M-dwarf stars in the *Kepler* sample, extrapolation suggests that the gap could potentially shift to ~1 $R_\oplus$ (e.g., like TRAPPIST-1 of ~ 0.1 solar mass ($M_\odot$)). The mass-radius (M-R) relation for Earth-like rocky planets (Zeng et al. 2016) is $(R_{planet}/R_\oplus) \approx (M_{planet}/M_\oplus)^{1/4}$. If the mass of planets scales with the mass of the star ($M_{planet} \propto M_{star}$), then, $R_{planet} \propto M_{star}^{1/4}$. Now, since the M-R relation for small stars ($M_{star}$ <1.66 $M_\odot$) is almost linear (Demircan & Kahraman 1991): $(R_{star}/R_\odot) \approx (M_{star}/M_\odot)$, it follows that $R_{planet} \propto R_{star}^{1/4}$. Thus, it is reasonable to expect a weak but non-trivial dependence of $(R_{gap}/2R_\oplus) \approx (R_{star}/R_\odot)^{1/4}$. This is consistent with the trend in the $R_{gap}$ in **Figure 1**. It can also explain why Fulton et al. (2017) claims there is no dependence on stellar radii, because it is a weak dependence (to the power of 1/4 of host star radii), and FGK stars all have similar radii. However, it makes important predictions for planets around M-dwarfs. This hypothesis can be tested by future observation of M-dwarf hosted planets.